\documentstyle[aps,pre,rotate,epsf]{revtex}
\begin{document}
\draft
\title{Comment on ``Analysis of chaotic motion and its shape dependence in a
generalized piecewise linear map''}
\author{R. Klages\cite{em}}
\address{Max Planck Institute for Physics of Complex Systems,
N\"othnitzer Str. 38,  D-01187 Dresden, Germany}
\date{\today}
\maketitle
\begin{abstract}
Rajagopalan and Sabir [Phys.\ Rev.\ E {\bf 63}, 057201 (2001)] recently
discussed deterministic diffusion in a piecewise linear map using an approach
developed by Fujisaka et al. We first show that they rederived the random walk
formula for the diffusion coefficient, which is known to be the exact result
for maps of Bernoulli type since the work of Fujisaka and Grossmann [Z. Physik
B {\bf 48}, 261 (1982)]. However, this correct solution is at variance to the
diffusion coefficient curve presented in their paper. Referring to another
existing approach based on Markov partitions, we answer the question posed by
the authors regarding solutions for more general parameter values by recalling
the finding of a fractal diffusion coefficient. We finally argue that their
model is not suitable for studying intermittent behavior, in contrast to what
was suggested in their paper.
\end{abstract}
\begin{figure}[b]
\epsfxsize=7cm
\centerline{\rotate[r]{\epsfbox{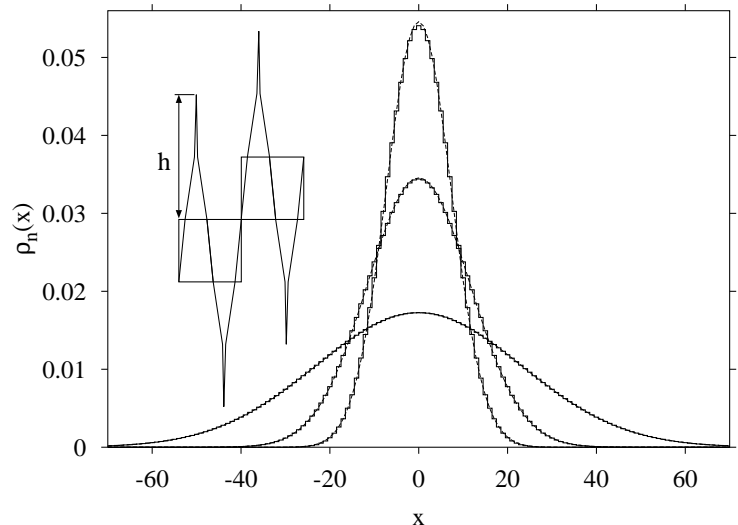}}}
\caption{Time-dependent probability density $\rho_n(x)$ for the map sketched
in the figure as it evolves starting from a uniform density in a box situated
around $x=0$. The results have been obtained from iterating transition
matrices as explained in the text. Included are Gaussian solutions from the
ordinary diffusion equation corresponding to the exact diffusion coefficient
$D(h,r)$ of the map, where $h=2, r=0.5$. These dashed lines are almost
indistinguishable from the map densities, however, they are lacking the
step-like fine structure. From above to below, the time steps are $n=20$,
$n=50$, $n=200$. The quantities plotted in this and in the following figures
are dimensionless.}
\label{fig1}
\end{figure}

The study of deterministic diffusion in simple chaotic maps on the line
appears to have originated about twenty years ago (see, e.g., Ref.\
\cite{RKdiss} and further references therein). Already in the seminal work by
Fujisaka and Grossmann \cite{GF2}, a variety of piecewise linear models was
defined and analyzed by means of stochastic modeling. All these maps are of
the form $x_{n+1}=M_h(x_n)$, where $h\in N$ is a control parameter, and $x_n$
is the position of a point particle at discrete time $n$. $M_h(x)$ is
continued periodically beyond the interval $[0,1)$ onto the real line by a
lift of degree one, $M_h(x+1)=M_h(x)+1$. The map defined in Ref.\
\cite{RaSa01}, which is sketched again in Fig.\ \ref{fig1}, provides a
straightforward generalization of the one introduced in Ref.\ \cite{GrTh83},
which is recovered at $h=1/2$. For this type of maps, indeed a vast literature
exists on how to obtain exact analytical results at specific cases of
parameter values; Refs.\ \cite{RKdiss,ddlit} sumarize some of these methods,
with more complete lists of references therein. It is furthermore well-known
that the calculations are particularly simple if the parameter is such that
the map exhibits the Bernoulli property \cite{Schu}.

\begin{figure}[b]
\epsfxsize=10cm
\centerline{\rotate[r]{\epsfbox{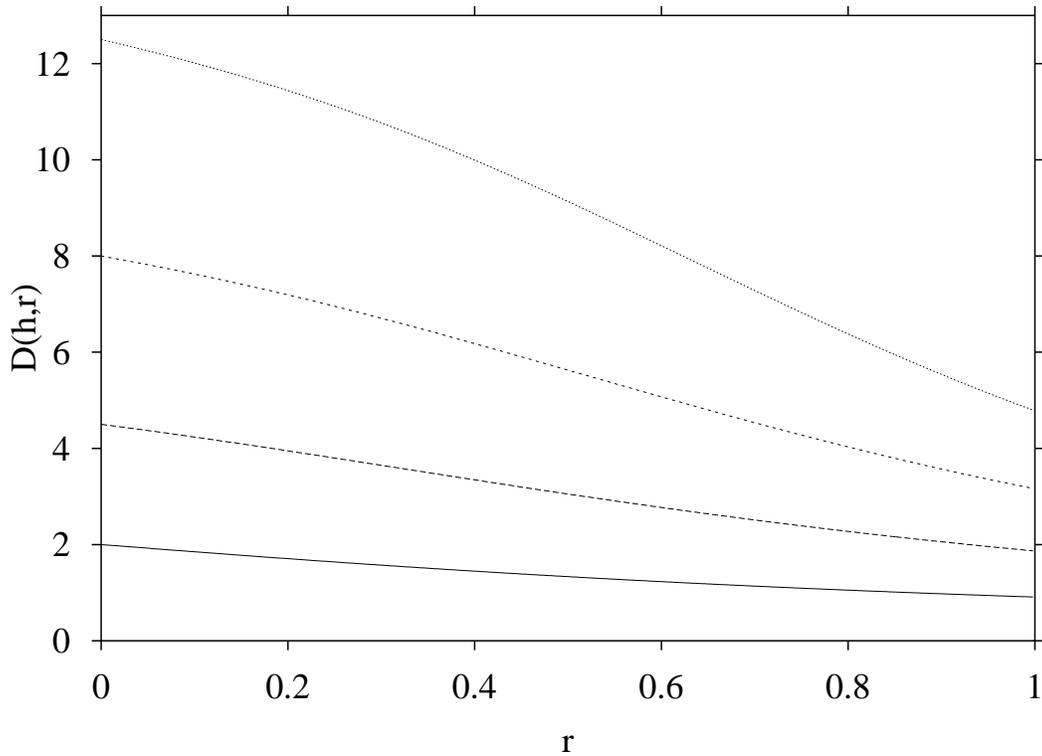}}}
\caption{The diffusion coefficient $D(h,r)$ for the map shown in Fig.\
\ref{fig1} according to Eqs.\ (\ref{eq:dk1}), (\ref{eq:dk2}). Solutions are shown for the
values of the height $h=2,3,4,5$ starting from below. The case $h=2$ corrects
the erroneous result in Fig.\ 3 of Ref.\ [3].}
\label{fig2}
\end{figure}

We first wish to present a considerable shortcut to the diffusion coefficient
calculations published in Ref.\ \cite{RaSa01}. Based on a theory which appears
to be a precursor of what was called ``Fujisaka's characteristic function
method'' in Ref. \cite{RaSa01}, Fujisaka and Grossmann have shown \cite{GF2}
that the diffusion coefficient formula
\begin{equation}
D=\frac{<j^2(x_n)>}{2} \label{eq:drw}
\end{equation}
provides the exact solution for types of maps as the one studied in Ref.\
\cite{RaSa01}, i.e., if they share the Bernoulli property. Here $j(x_n)$ is
the jump velocity defined as $j(x_n):=[x_{n+1}]-[x_n]$ with $[x]$ being the
largest integer less than $x$, and $<\ldots>$ denotes the average over the
invariant probability density. This expression is just identical to the
familiar random walk formula for diffusion on a one-dimensional lattice, where
the length of jumps squared is weighted with the probability to perform such
jumps (see Ref.\ \cite{dcrc} and further references therein). Evaluating this
equation for the map under consideration leads to
\begin{equation}
D(h,r)=\sum_{j=1}^hj^2\delta(j,r) \label{eq:dk1}
\end{equation}
with $h\in N , 0<r<1$, where $2\delta(j,r)$ denotes the probability
to jump over a distance of $j$ steps and is easily calculated to
\begin{equation}
\delta(j,r)=\frac{2}{m_j} \;, \label{eq:dk2}
\end{equation}
$m_j$ being the slopes of the map. Combining the above two equations yields
Eq.\ (17) of Ref.\ \cite{RaSa01}. We conclude that Rajagopalan and Sabir have
confirmed again Eq.\ (\ref{eq:drw}) of Fujisaka and Grossmann as applied to
their specific map. We now focus on the author's special case of the map
defined by the relation for the slopes
\begin{equation}
m_0=3+\frac{4(1-r^h)}{r^h(1-r)}
\end{equation}
with $m_i/m_{i-1}=r$. Solutions for Eqs.\ (\ref{eq:dk1}), (\ref{eq:dk2}) under
this constraint are shown in Fig.\ \ref{fig2} for different $h$. This figure
corrects the erroneous result shown in Fig.\ 3 of Ref.\ \cite{RaSa01}, which
only includes a few data points and appears to indicate a rather irregular
curve for the diffusion coefficient at $h=2$. Below we will explain why all
the curves shown in Fig.\ \ref{fig2} must indeed be simple functions of $r$.

However, first we would like to recall a second method which is not restricted
to special cases of parameters such as integer heights, in contrast to the one
outlined in Refs.\ \cite{GF2,RaSa01}. The basic idea of this method is to
directly solve the Frobenius-Perron equation of the dynamical system,
\begin{equation}
\rho_{n+1}(x) = \int dy \; \rho_n(y) \; \delta(x-M_h(y)) \; ,
\end{equation}
where $\rho_n(x)$ is the probability density for points on the real
line. There exists a dense set of parameter values $h$ for which one can
construct Markov partitions of the map, and for each of these parameter values
this equation can be written as a matrix equation
\cite{RKdiss,RKD},
\begin{equation}
\mbox{\boldmath $\rho$}_{n+1}=\, T(h,r) \, \mbox{\boldmath
$\rho$}_n\;.
\end{equation}
$\mbox{\boldmath $\rho$}_n$ represents a column vector of the probability
densities defined on each part of the Markov partition at time $n$, and
$T(h,r)$ is a topological transition matrix which can be constructed from the
Markov partition. This setup provides two ways of solution: one way is to
solve the eigenvalue problem of $T(h,r)$ and to relate the diffusion
coefficient to its eigenvalues. As is shown in detail in Refs.\ \cite{RKD},
in special cases all calculations can be performed analytically. For the
simple map defined in Ref.\ \cite{RaSa01} these calculations are
straightforward and confirm again Eqs.\ (\ref{eq:dk1}), (\ref{eq:dk2}).  For
more general cases, the matrix equation can simply be iterated
\cite{RKdiss,dcrc} yielding numerically exact solutions for the probability
density vector $\mbox{\boldmath $\rho$}_n$ at any time step $n$, as well as
for any other dynamical quantity based on probability density averages. Both
such methods were previously applied to various examples of piecewise linear
maps \cite{RKdiss,dcrc,RKD}. Fig.\ \ref{fig1} presents analogous
results for the map studied in Ref.\ \cite{RaSa01} at $h=2$ and $r=0.5$, cp.\
to Fig.\ 3.1 on p.54 of Ref.\ \cite{RKdiss}. The probability density is a
Gaussian on a coarse scale, whereas the fine scale is determined by the
invariant density of the map on the unit interval. These deviations from an
exact Gaussian can quantitatively be evaluated, e.g., by calculating the
curtosis of the respective map density; for a more detailed discussion of such
aspects we refer to Chapter 3 of Ref.\ \cite{RKdiss}. This interplay between
fine and coarse structure of the probability densities was furthermore
discussed in terms of the spectrum of eigenmodes of the Frobenius-Perron
operator, see \cite{RKdiss,RKD}.  These known results appear to be
recovered in Ref.\ \cite{RaSa01} by a respective analysis of the fluctuation
spectrum, which provides an alternative way to look at the probability density
of the map. 

\begin{figure}[b]
\epsfxsize=10cm
\centerline{\rotate[r]{\epsfbox{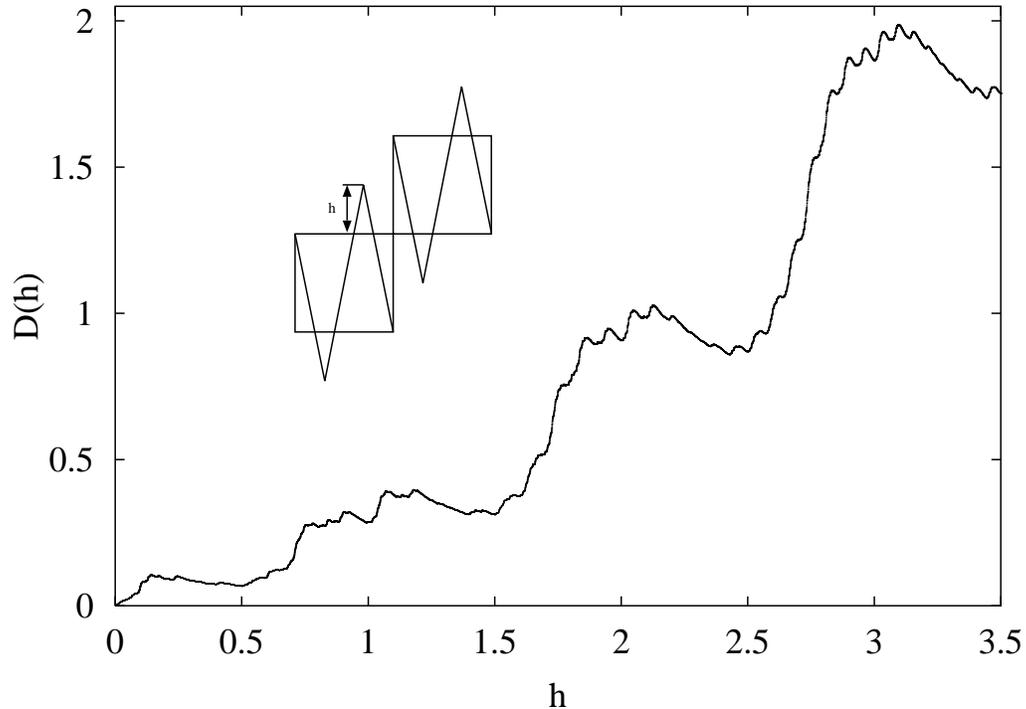}}}
\caption{Fractal diffusion coefficient $D(h)$ for the mirrored zigzag-map
sketched in the figure as a function of the height $h$. Shown are 13376 data
points. The data is from Refs.\ [1,5].}
\label{fig3}
\end{figure}

In their outlook to further work, the authors of Ref.\ \cite{RaSa01} raised
the question of how to compute the diffusion coefficient for maps with
fractional heights $h$, and how it may look like. The application of the
arsenal of methods outlined above has already given a full answer to this
problem. As a central result, it was found that the diffusion coefficient for
these maps is a fractal function of the parameter $h$. To present an example,
Fig.\ \ref{fig3} depicts the result for a mirrored zigzag-map with uniform
slope, which has some similarities with the one studied in Ref.\
\cite{RaSa01}; for details see Refs.\ \cite{RKdiss,dcrc}. Knowing these
results, it is straightforward to conclude that, for arbitrary height $h$, the
map studied in Ref.\ \cite{RaSa01} will just yield another fractal diffusion
coefficient; further evidence for that statement is provided by the numerical
and analytical data presented in Ref.\ \cite{Tse94}.

So why can the diffusion coefficient of the map in Ref.\ \cite{RaSa01} not be
fractal as a function of $r$ at integer values of the height? One way to look
at this problem is to inquire how the topology of the map is affected by
parameter variation. A fundamental tool providing detailed information about
the topology of a dynamical system are Markov partitions. Varying $r$ at
integer heights does not change the Markov partition, thus the topology of the
map does not change, and any quantity resulting from an average over the
invariant density is a simple function of the parameter
\cite{StSt97}. However, changing the height changes the Markov partition in a
complicated way and reflects the topological instability of the map under this
type of parameter variation. This topological instability results in fractal
transport coefficients.

Finally, we comment on the conclusion of Rajagopalan and Sabir that the map
studied in their paper is ``suited in describing diffusion systems showing
intermittency''. In this aspect the authors appear to follow Ref.\
\cite{GrTh83}, where the map shown in Fig.\ \ref{fig1} at $h=1/2$ was
introduced for the purpose of modeling ``strong correlations between
successive steps... as realized in Brownian motion with directional
persistence''. Indeed, Grossmann and Thomae revealed a persistent dynamics
which they characterized as ``intermittent-{\em like}'' behavior. They linked
these correlations to deviations from a pure Gaussian probability density such
as the ones discussed above.

In the following we use the term ``intermittency'' in the sense of Pomeau and
Manneville (see, e.g., Ref.\ \cite{Schu} for a tutorial about their
results). Particularly, we wish to distinguish it from the denotation
``intermittency-like'' in the sense of Grossmann and Thomae. Extensive studies
of diffusion in one-dimensional intermittent maps led to the conclusion that,
generally, in this case a diffusion coefficient does not exist
\cite{anom}. Furthermore, all maps studied in these references are inherently
nonlinear. For piecewise linear expanding maps which are uniquely ergodic if
restricted to compact spaces, such as the one of Refs.\ \cite{RaSa01,GrTh83},
there is no evidence for intermittency nor for anomalous diffusion
\cite{RKdiss}. Applying the concept of conjugacy enables to transform
piecewise linear maps onto nonlinear ones. However, the diffusive dynamics is
invariant under conjugacy \cite{GrTh83}, thus the corresponding nonlinear map
is again non-intermittent and normal diffusive. To our knowledge the only
piecewise linear map exhibiting intermittency was introduced in Ref.\
\cite{GaWa88}, and it belongs to a very different class than the one of Refs.\
\cite{RaSa01,GrTh83}.

In summary, by relating the piecewise linear map studied in Ref.\
\cite{RaSa01} to intermittent behavior the authors confuse the meaning of
intermittency, in the sense of Pomeau and Manneville, with the existence of
intermittent-{\em like} behavior, in the sense of persistence in the diffusive
motion. Intermittency generally leads to anomalous diffusion, whereas
persistence in piecewise linear maps shows up in form of local extrema of the
fractal diffusion coefficient at integer and half-integer heights, see Fig.\
\ref{fig3}. We conclude that the analysis of chaotic motion and its shape
dependence as performed in Ref.\ \cite{RaSa01} has nothing to do with
intermittency, but instead recovers features of the parameter-dependent normal
diffusion coefficient as studied in Refs.\
\cite{RKdiss,GF2,GrTh83,ddlit,Schu,dcrc,RKD,Tse94,StSt97}.

The author thanks N.Korabel and J.R.Dorfman for helpful remarks.

\end{document}